\documentclass{article}
\usepackage{srcltx}
\usepackage[T2A]{fontenc}
\usepackage[T1]{fontenc}
\usepackage[cp1251]{inputenc}
\usepackage{graphicx}
\usepackage{latexsym}
\usepackage{amscd}
\usepackage{amsmath}
\usepackage{amssymb}
\usepackage{indentfirst}
\usepackage[small,centerlast]{caption2}
\usepackage{longtable}
\usepackage[dvipsnames]{color}
\usepackage{cite}

\begin{document}

\title {APPROXIMATE ANALYTIC MODEL OF THE BOUNDARY LAYER AROUND A LOW MAGNETIC FIELD NEUTRON STAR AT THE DISK ACCRETION}

\author{Bisnovatyi-Kogan G. S.
\footnote{Space Research Institute RAS, Moscow, RF}
\footnote{MEPhI, Moscow, RF}}
\date{}
 \maketitle

\begin{abstract}

\noindent An approximate analytic one-dimensional model 
is constructed, for the accretion disk boundary
layer surrounding a neutron star whose low magnetic field does not affect the process of accretion. A high luminosity model is considered, with radiation pressure dominant in the interior part of the disk.
\end{abstract}


\section{Introduction}

Accretion onto neutron stars (NSs) and black holes (BHs) is what predominantly supplies galactic X-ray
sources with radiation energy. Observational evidence indicates that these objects are, in the vast majority of cases,
close binary systems, where mass is transferred from a stellar companion onto the NS or BH \cite{sht85,lip87}. Due to high
angular momentum of the accreting material in such systems, accretion proceeds in the form of a disk. In the case
of black holes, the material in the disk reaches the radius of the innermost stable circular orbit and then falls nearly
freely onto the BH. The main characteristics of the disk accretion onto BHs are given in approximate models
developed in \cite{sha,shs,nt,bkb,pbk}. The outer regions of accretion disks around NSs are generally modeled similarly to those around
BHs. Significant differences arise, however, when the accretion disk is truncated by the strong magnetic field of a
neutron star \cite{bkfr,lip87}, or when the disk extends down to the surface of a NS with a weak magnetic field. In the latter
case, a boundary layer (BL) forms near the surface of the neutron star, where the rotational velocity of the accreting
matter rapidly decreases from the Keplerian angular velocity in the disk to the much lower angular velocity of the
NS. This deceleration occurs within the BL, whose radial thickness is smaller than its vertical extent.
The rapid decrease in angular velocity and the small radial thickness of the boundary layer (BL) surrounding
a neutron star (NS) pose significant challenges for the development of computational models of such a BL that remain
unresolved to this day. A number of studies have been devoted to constructing approximate analytical models of the
BL around NSs, employing various simplifying assumptions. The resulting models sometimes yield significantly
different outcomes \cite{pring1,pring2,tyl1,tyl2,reg,pap,ss,pop,is}.
In the present work, we construct an approximate model of the BL around a weakly magnetized, high-luminosity
NS, when pressure is determined by radiation in both the inner regions of the accretion disk and within
the BL itself. Under these conditions, the expressions for pressure $P$, specific energy $E$, and entropy $S$ are
approximated as follows:

 \begin{equation}
P=\frac{aT^4}{3},\quad E=\frac{aT^4}{\rho},\quad S=\frac{4aT^3}{3\rho}.
 \label{rdom}
\end{equation} 
In obtaining the approximate solution for the accretion disk with a boundary layer (BL), coarser simplifying
assumptions were employed compared to previous studies. This approach helped us to obtain analytical expressions
for the distribution of $P,\,\,\rho,\,\,T,\,\,\omega,\,\,h$ within the BL. The matching of the solutions for the disk and the BL was
performed at the point where the distance from the neutron star surface is equal to the vertical thickness of the BL
at that location. We discuss also the process of boundary layer disappearance as the rotational velocity of
neutron star approaches the critical value, at which the gravitational attraction at the NS surface is balanced by the
centrifugal force from the Keplerian rotation.

\section{One-dimensional model of accretion disk}

Let us consider a one-dimensional model \cite{sha,shs,nt} that approximately accounts the dependence of variables along
the vertical $z$-axis. This model also employs an approximate expression for the viscous component of the energy-momentum
tensor in a turbulent disk, given by:

\begin{equation}
t_{r\phi}=-\alpha \rho v_s^2=-\alpha P.
 \label{tvis}
\end{equation}
The square of the mean turbulent velocity is proportional to the square of the sound speed $v_t^2=\alpha v_s^2=\alpha\frac{P}{\rho}$, with the parameter  $\alpha < 1$ in the subsonic turbulence. The approximate model is using an isothermal sound speed,
neglecting the dependence on the adiabatic index. The same viscosity prescription is used for both the disk and the
boundary layer. In this approximate one-dimensional model of a thin disk, Keplerian angular velocity is assumed,
and the disk is considered as a stationary one. Under these assumptions, the inward angular momentum flux toward the neutron
star due to disk accretion is balanced by the outward flux due to turbulent viscosity:

\begin{eqnarray}
\Omega=\Omega_K=\left(\frac{GM}{r^3}\right)^{1/2},\quad h\approx
\frac{v_s}{\Omega_K},\qquad  \nonumber \\
\dot M (j-j_{in})= -2\pi r^2 2h t_{r\phi}, \quad  t_{r\phi}=
\eta r \frac{d\Omega}{dr}=-\alpha P.
  \label{1dd}
\end{eqnarray}
Here, $\dot M$ is the prescribed mass accretion rate,
 $j=v_\phi r=\Omega r^2$ is the specific angular momentum of the matter in the
disk, and $j_{in}$ refers to the angular momentum of matter around the neutron star with a radius $r_*$, to what it is 
accreted. The disk half-thickness $h(r)$ is approximately determined from the vertical hydrostatic equilibrium equation
along the $z$-axis. Both expressions for $t_{r\varphi}$ in equation ({\ref{1dd}) coincide when $\eta=\frac{2}{3}\frac{\alpha P}{\Omega_K}$, but the computational
methods used for numerical modeling differ significantly.

To approximately determine the structure of a steady-state one-dimensional accretion disk, the
 equi\-librium
equations presented above must be supplemented with equations of thermal equilibrium. In the model developed in
\cite{sha}, it was assumed that all the heat generated by viscous dissipation $Q_+$, is radiated away through the disk surface
via radiative heat transfer, so that $Q_+=Q_-$. We have than:

\begin{eqnarray}
Q_+=h t_{r_\phi} r \frac{d\Omega}{dr}=\frac{3}{8\pi}\dot M \frac{GM}{r^3}
\left(1-\frac{j_{in}}{j}\right),\quad Q_-\approx \frac{4}{3}\frac{acT_0^4}{\varkappa \Sigma_0},\nonumber\\
 \Sigma_0=2\rho_0 h, \quad v_r\approx -\frac{\dot M}{4\pi r h \rho_0}=
 -\frac{\dot M}{2\pi r\Sigma_0}, \quad \dot M>0,\quad v_r<0 \qquad.
 \label{1dt}
\end{eqnarray}
Here,  $\varkappa$  is the opacity, and  $\rho_0$ $T_0$ are the pressure and the temperature of the disk in its equatorial plane.
Under the condition $Q_+=Q_-$, it follows from equation ({\ref{1dt}) that:

\begin{equation}
\frac{3}{2}{\dot M} \Omega^2 \left(1-\frac{j_{\rm in}}{j}\right) =
   \frac{16 \pi acT_0^4}{3\varkappa \Sigma}\,.
\label{eq:11.P2.06}
\end{equation}
\noindent
In the central region of the accretion disk around a black hole, the pressure is dominated by radiation, $P_r=\frac{aT^4}{3}$,
and the opacity is primarily due to scattering on electrons, $\varkappa=\varkappa_T=0.4$ cm$^2$/g (for a hydrogen plasma). In this case,
from equations (\ref{1dd})-(\ref{eq:11.P2.06}), we obtain, for the radial dependencies at $r>r_*$, the following expressions for
the disk half-thickness $h(r)$, pressure $P_0(r)$, density $\rho_0(r)$, temperature $T_0(r)$, radial velocity in the equatorial plane
$v_{r0}$, and surface density $\Sigma_0(r)$, \cite{bk2010}:

\[
h=\frac{\varkappa_T}{c}\frac{Q_+ r^3}{GM}=
   \frac{3}{8 \pi}\frac{\varkappa_T}{c}{\dot M} \left(1-\frac{j_{\rm in}}{j}\right)
\]
\begin{equation}
\quad =\frac{3GM_{\odot}}{2c^2} m\, {\dot m}\left(1-\frac{j_{\rm in}}{j}\right)=
   7.41\times 10^4m\, {\dot m}\left(1-\frac{j_{\rm in}}{j}\right)\,.
\label{eq:11.P2.08}
\end{equation}

\begin{equation}
P_0=\frac{2c \Omega}{3 \alpha \varkappa_T}=\frac{}{3}\frac{c^4}{
  \alpha \varkappa_T GM_{\odot}}\frac{1}{mx^{3/2}}=\frac{1.01 \times 10^{16}
 }{\alpha m x^{3/2}}\,.
\label{eq:11.P2.09}
\end{equation}

\[
\rho_0=\frac{\Sigma}{2h}=\frac{256 \pi^2}{27 \alpha}\frac{c^3}{\varkappa_T^3}
 \frac{1}{\Omega{\dot M}^2} \left(1-\frac{j_{\rm in}}{j}\right)^{-2}
\]
\begin{equation}
\quad =\frac{16}{27\alpha}\frac{c^2}{\varkappa_T GM_{\odot}}
  \frac{x^{3/2}}{{\dot m}^2 m} \left(1-\frac{j_{\rm in}}{ j}\right)^{-2}=
 \frac {10^{-5}}{\alpha}
 \frac{x^{3/2}}{{\dot m}^2 m} \left(1-\frac{j_{\rm in}}{j}\right)^{-2}.
 \label{eq:11.P2.11}
\end{equation}

\begin{equation}
T_0=\left(\frac{ 3P }{ a} \right)^{1/4}=\left(\frac{2 }{ \alpha\varkappa_T aGM_{\odot}}
   \right)^{1/4}\frac{c }{ m^{1/4}x^{3/8}}=\frac{4.48 \times 10^7 }{
   m^{1/4}x^{3/8}}\,\,{\rm K}\,.
 \label{eq:11.P2.13}
\end{equation}

\[
v_{r0}=-\frac{{\dot M} }{ 2\pi r \Sigma}=-\frac{9\alpha }{ 128 \pi^2}
  \frac{\varkappa_T^2 }{ c^2}\frac{\Omega }{ r}
  {\dot M}^2 \left(1-\frac{j_{\rm in}}{ j}\right)
\]
\begin{equation}
\quad=-\frac{9\alpha }{8}c\frac{{\dot m}^2 }{ x^{5/2}}
  \left(1-\frac{j_{\rm in}}{ j}\right)\,.
 \label{eq:11.P2.12}
\end{equation}

\[
 \Sigma= \frac{64 \pi }{ 9 \alpha}\frac{c^2 }{\varkappa_T^2}\frac{1 }{ \Omega
  {\dot M}} \left(1-\frac{j_{\rm in}}{ j}\right)^{-1}\]
\begin{equation}
\quad =\frac{16 }{ 9\alpha \varkappa_T}
  \frac{x^{3/2} }{ {\dot m}} \left(1-\frac{j_{\rm in}}{ j}\right)^{-1}  =
  \frac{40 }{ 9\alpha}\frac{x^{3/2} }{ {\dot m}}
  \left(1-\frac{j_{\rm in}}{ j}\right)^{-1}\,,
\label{eq:11.P2.10}
\end{equation}
\noindent
and also

\begin{equation}
\vert\frac{v_r }{ v_s}\vert=
  \frac{3\sqrt2\,\alpha }{ 16 \pi}\frac{\varkappa_T{\dot M} }{ cr}=
  \frac{3\sqrt2 \, \alpha }{ 4} \frac{{\dot m} }{ x}.
 \label{eq:11.P2.14}
\end{equation}

\noindent
With appropriate corrections in the region adjacent to the boundary layer (BL), this solution can be applied to model
accretion onto a neutron star (NS). All numerical quantities are expressed in CGSE units, with $r=r_*$  denoting the
radius of the neutron star, and the following notations have been used:

\begin{equation}
m=\frac{M }{ M_{\odot}},\quad {\dot m}=\frac{{\dot M}c^2 }{ L_c}, \quad
   L_c=\frac{4\pi cGM }{ \sigma_T}, \quad x=\frac{rc^2 }{ GM}.
 \label{eq:11.P2.15}
\end{equation}
\noindent
\[
 v_s^2=\frac{P }{ \rho}, \quad \Omega=\Omega_K=\frac{c^3 }{ GM}x^{-3/2}=
   \frac{2.02 \times 10^5 }{ mx^{3/2}}, \quad
   j_{in}=\xi j_{K*}=\xi \Omega_{K*}r_*^2, \quad \xi \leq 1.
\]
The angular momentum acquired by the neutron star during steady-state accretion corresponds to the value
at the point of the maximum angular velocity within the boundary layer (BL). Due to the small radial thickness of
the BL, the radius at all points within it is taken to be equal to $r_*$. In what follows, we assume that the entire angular
momentum flux of the matter accreting onto the neutron star is absorbed by it. The angular momentum of the infalling
matter is determined by the Keplerian rotational velocity at the neutron star surface, which defines the parameter
 $\xi=1$.

\section{ One-Dimensional Model of the Boundary Layer}

Let us consider the equations describing the boundary layer, in which the angular velocity decreases as the
flow approaches the neutron star. The radius extent of the BL is so small that in all algebraic expressions not involving
differentiation, it was used $r=r_*$. The vertically averaged, approximate equations of motion have been written in
the form  \cite{tyl2,reg}:

\begin{equation}
\label{4.4}
\frac{d P_0 }{ dr}=-\Omega_{K \ast}^2 r_* \rho_0
(1-\omega^2),
\end{equation}

\begin{equation}
\label{4.5}
\frac{d\omega }{ dr}=\frac{\dot M }{  4 \pi h \eta_b r_{\ast}}
(1-\omega)=\frac{\dot M \Omega_{K \ast}}{ 4 \pi r_* \alpha_b}\,
\frac{1-\omega}{\rho_0 v_s^2 H_b},
\end{equation}
where
\begin{equation}
\label{4.6}
\omega = \frac{\Omega }{ \Omega_{K \ast}},\quad \Omega_{K \ast}^2=
\frac{GM }{ r_{\ast}^3}.
\end{equation}

Here, only the principal terms are retained in the equation of motion;  $j_{in}=\Omega_{K \ast}r_*^2$, so that the angular momentum
flux onto the neutron star corresponds to expression (\ref{eq:11.P2.15}). Except calculations of derivatives,  everywhere within the boundary layer, the radius  $r$  is taken equal to $r_*$ \cite{reg,bkl2001}. In the accretion disk, the turbulent viscosity coefficient was modelled as:

\begin{equation}
\label{4.6a}
\eta=\frac{2}{3}\frac{\alpha P}{\Omega}\approx \frac{2}{3}\alpha\rho\frac{v_s^2}{\Omega}\approx \frac{2}{3}\alpha\rho v_s h.
\end{equation}
Here the disk half-thickness $h$ is taken as the minimal scale. Inside the boundary layer, the minimal scale is its radial thickness;
therefore, the BL turbulent viscosity coefficient is considerably smaller and is assumed in the form

\begin{equation}
\label{4.6b}
\eta_b\simeq \alpha_b\rho v_s H_b,
\end{equation}
at $\alpha,\,\,\alpha_b \lesssim 1$. 

 \section{Thermal Processes in the Boundary Layer}

The rapid decrease of angular velocity within the boundary layer during accretion onto a slowly rotating
neutron star results in strong heat production due to viscous dissipation, and intense heating of the matter in the
boundary layer. In a thin boundary layer, the radial heat flux exceeds the heat losses through the disk surfaces. Taking
into account the angular momentum conservation equation (\ref{1dd}), the expression for the viscous heating rate of the
boundary layer half-thickness $Q_+$ (erg/cm$^2$/s), with account of relations (\ref{eq:11.P2.15}),(\ref{4.5}), can be written as:

\begin{equation}
\label{blt1}
Q_+=h t_{r\phi}r \frac{d\Omega}{dr}
=h\eta \left(r\frac{d\Omega}{dr}\right)^2
\approx -\frac{GM{\dot M}}{4\pi r_*^2}(\omega-1)\frac{d\omega}{dr}.
\end{equation}
Inside the boundary layer, it is convenient to introduce the quantity $Q_t$ (in erg/cm/s)

\begin{equation}
\label{blt2}
Q_t=4\pi r Q_+ \approx 4\pi r_* Q_+
= -\frac{GM{\dot M}}{r_*}(\omega-1)\frac{d\omega}{dr},
\end{equation}
defining  heating rate due to viscous dissipation in the ring of unit radial thickness. Along the radius, heat transfer occurs through radiative
conductivity, so the radial flux in the disk  $\Phi$ (erg/s) is written as

\begin{equation}
\label{blt3}
\Phi=-\frac{4ac T^3}{3\kappa \rho}\frac{dT}{dr}\,4\pi r_* h.
\end{equation}
The equation of heat balance within the boundary layer has the form

\begin{equation}
\label{blt4}
\frac{d\Phi}{dr}=Q_t
=-\frac{GM{\dot M}}{r_*}(\omega-1)\frac{d\omega}{dr}.
\end{equation}
Integrating this equation with the boundary condition $\Phi(\omega_*)=0$, at which all the heat flux is directed along the disk, outside of the neutron star, we obtain:

\begin{equation}
\label{blt5}
\Phi=\frac{GM{\dot M}}{r_*}\left(\omega-\frac{\omega^2}{2}
-\omega_* + \frac{\omega_*^2}{2}\right)
=\frac{GM{\dot M}}{2 r_*}(\omega-\omega_*)(2-\omega-\omega_*).
\end{equation}
Using here $\omega=1$, we obtain the value of the viscous heat flux exiting the boundary layer,$\Phi_{b}$ (erg/s), in the form
\cite{pap}:

\begin{equation}
\label{blt6}
\Phi_{b}= \frac{GM{\dot M}}{2r_*}(1-\omega_*)^2.
\end{equation}
The whole thermal flux  here is directed outside the neuron star, 
what happens when the neutron star is sufficiently hot. During accretion into the cold neutron star, part of the heat flux   
is used for its heating.

\section{Global Model of the Accretion Disk with a Boundary Layer}

To describe the boundary layer (BL), the equations of motion in the form (\ref{4.4}),(\ref{4.5}) were used, where
the $t_{r\varphi}$ component was defined by the angular velocity gradient $\frac{d\Omega}{dr}$, and the viscosity coefficient 
$\eta_b$ from (\ref{4.6b}).
In the paper \cite{pop}, the formula was employed with a continuous decrease of the viscosity coefficient when transitioning
from the disk to the BL due to the reduction of the characteristic scale from $h$  in the disk to $H_b$ in the BL, which
in  \cite{bkl2001} was written as

\begin{equation}
\label{blt8}
\eta\, = \,
\frac{2\rho_0 v_s}{3}\left[\frac{1}{(\alpha \,h)^2}+
\frac{(2\,dP_0/dr)^2}{(3\alpha_b\, P_0)^2}\right]^{-1/2},
\end{equation}
to ensure consistency with the limiting values (\ref{4.6a}) for the disk and (\ref{4.6b})  for the BL, respectively. At the same time,
$h \ll H_b \approx \frac{P_0}{dP_0/dr}$ in the disk, and 
$h \gg H_b$ in the BL.
In the thermal balance equation, the radiative heat flux $\Phi$ along the radius from $\Phi$ should be taken into
account, as it exceeds the heat transfer along the disk axis. In the boundary layer, the angular velocity can
significantly differ from the Keplerian value; therefore, the target equation should be expressed through the gradient
of the local angular velocity. Considering the advective heat transport along the radius \cite{pbk}

\begin{equation}
\label{blt8a}
Q_{adv}=-\frac{\dot M}{2\pi r} T\frac{dS}{dr},
\end{equation}
(S is entropy), the equation of energy balance within the boundary layer, with local viscous heating $Q+$, and radiative
radial flux  $Q_-$, is written as \cite{bkl2001}

\begin{equation}
\label{blt9}
Q_+\,=\,Q_-\,+\,Q_{adv}\,+\,\frac{1}{4\pi r}\frac{d\Phi}{dr}.
\end{equation}
Here, $\Phi$ from the  relation (\ref{blt3}) is used. Substituting the expressions for $Q+$ and $Q_-$, we obtain this equation in
the form

\begin{equation}
\label{blt10}
{\dot M}\,T \frac{dS}{dr}\,=
\frac{d\Phi}{dr}\,+\,{\dot M}(j-j_{\rm in})\frac{d\Omega}{dr}
+\,\frac{2aT_0^4c}{3\varkappa \rho_0 h} 4\pi r.
\end{equation}
The disk thickness, as well as in the boundary layer, is approximately determined by (\ref{1dd}) from the vertical
hydrostatic equilibrium equation. In \cite{pop}, a numerical solution of the equations for the disk with a boundary layer
surrounding a white dwarf was obtained under boundary conditions at the stellar radius $r_s$. Due to the no-slip boundary
condition, the angular velocity at the BL inner boundary was equal to the stellar rotation velocity, $\Omega_b(r_s)=\Omega_s$. At
the outer boundary of the disk, taken at $r_{out}=100r_*$, the Keplerian velocity $\Omega_{out}=\Omega_K(r_{out})$ was assumed.

We consider the boundary condition for the viscous angular momentum transport as free inflow and absorption
by the star, neglecting heating of the neutron star through viscous dissipation. Numerical solutions of the global equations
for the disk and the boundary layer were obtained in \cite{pop} for cataclysmic variables with a boundary layer, where the
accretion disk meets the accreting white dwarf. A similar global numerical solution for a neutron star has not yet been
obtained.

The construction of a global model for accretion onto a neutron star with a boundary layer was carried out
in  \cite{reg} using the method of matched asymptotic expansions (MAE) \cite{nay81}. Obtaining a direct numerical solution is
complicated by the presence of a very small parameter representing the ratio of the layer thickness to the radius and to the
thickness of the accretion disk. A solution accounting for thermal processes was numerically constructed in \cite{reg}, and was characterized by
a physically unjustified flux of the outward angular momentum transport along the disk radius. This may
indicate poor applicability of the MAE method in this model\cite{bkl2001}. For a polytropic model of the accretion disk with
a boundary layer, a global solution using the MAE method was constructed in \cite{bk94}.

\section{Analytical Model of the Accretion Disk and Boundary Layer around a Neutron Star}

Considering both the equation of state $P(\rho, T)$ and thermal processes in the disk has made the problem so
complex that, to date, even an approximate analytical solution has not been achieved. We have derived an approximate
analytical solution for the disk and boundary layer model in an important special case of higher luminosity,
where radiation pressure dominates in the interior part of the disk. We employed the method of matching the exterior
(disk) and interior (boundary layer) solutions, with boundary conditions differing from those used in \cite{reg},\cite{pop}, along
with a change of variables for finding the interior solution in the boundary layer.

As the equilibrium solution for the accretion disk, where the boundary layer has no discernible influence, the
standard approximate solution with $\Omega=\Omega_K$, given in (\ref{eq:11.P2.08})-(\ref{eq:11.P2.12}), was used.

As follows from (\ref{blt1}), near the neutron star radius, where the angular velocity reaches its maximum$\omega\approx 1$,
viscous dissipation vanishes and viscous angular momentum transport changes direction. In the steady state, the
angular momentum flux at the neutron star is equal to 
$j_{\rm in}\approx \Omega_{K*}\,r_*^2$. The angular momentum flux at the neutron star
sharply decreases upon reaching the maximum angular velocity 
$\Omega_{*c}$ of the uniformly rotating neutron star. Correspondingly,
the outward angular momentum flux along the accretion disk and its luminosity increase \cite{bk93}.

Viscous heat generation is very significant in the boundary layer due to the large angular velocity gradient. The thermal balance equations (\ref{blt10}),(\ref{blt3}) are solved jointly with the radial equilibrium equations and the angular
momentum balance equation (\ref{4.4})- (\ref{4.6}). Over the small boundary layer thickness $H_b$, the parameters $T,\, \rho,\, P,\, \omega$ vary strongly.
We use equation (\ref{4.5}) to replace the independent variable $r$ with $\omega$, so that

\begin{equation}
\label{blt12}
dr=\frac{4 \pi r_* \alpha_b }{ \dot M \Omega_{K \ast}}\,
\frac{\rho_0 v_s^2 H_b}{1-\omega} d\omega  ,
\end{equation}

Considering \ref{blt12}), the equation (\ref{4.4}) can be rewritten as,

\begin{equation}
\label{4.4b}
\frac{d P_0 }{ d\omega}=-\Omega_{K \ast}^2 r_* \rho_0(1-\omega^2)\frac{4 \pi r_* \alpha_b }{ \dot M \Omega_{K \ast}}\,
\frac{\rho_0 v_s^2 H_b}{1-\omega}
=-\Omega_{K \ast}^2 r_* \rho_0(1+\omega)\frac{4 \pi r_* \alpha_b }{ \dot M \Omega_{K \ast}}\,
\rho_0 v_s^2 H_b.
\end{equation}
The approximation used here is not applicable for neutron star rotation speeds near $\Omega_{*c}$, when the neutron
star becomes oblate due to the strong centrifugal force, the BL thickness becomes comparable to the accretion disk
thickness, and the boundary layer disappears upon reaching
$\Omega_{*}=\Omega_{*c}$  \cite{bk93}.

For moderate rotation velocities, when deviations from sphericity can be neglected, the boundary layer
thickness $H_b$ is taken from  (\ref{blt8}) as a characteristic length scale of pressure variation \cite{pop}. Using, by definition, the
expression for pressure $P_0=\rho_0 v_s^2$, we obtain from  (\ref{4.4b}),(\ref{4.4})

\begin{equation}
\label{4.4c}
\frac{d P_0 }{ dr}=-\Omega_{K \ast}^2 r_* \frac{P_0}{v_s^2}  
(1-\omega^2),\quad
{H_b}=\mid{\frac{P}{dP/dr}}\mid\approx\frac{v_s^2}{\Omega_{K \ast}^2 r_*(1-\omega^2)}=
h\frac{v_s}{\Omega_{K \ast} r_*(1-\omega^2)}=\frac{h^2}{r_*(1-\omega^2)},
\end{equation}
\begin{equation}
\label{4.4d}
\frac{d P_0 }{ d\omega}=-
\Omega_{K \ast}^2 r_* \rho_0\frac{4 \pi r_* \alpha_b }{ \dot M \Omega_{K \ast}}\,
P_0 \frac{v_s^2}{r_*\Omega_{K \ast}^2(1-\omega)},\quad h=\frac{v_s}{\Omega_{K*}}.
\end{equation}
Here, the expression for $H_b$ is applicable for 
$H_b<h$,\,\,\, $\omega^2< 1-\frac{v_s}{r_*\Omega_{K \ast}}=\omega_b^2$. The point where $h=H_b$ can be
assumed as the boundary separating the accretion disk from the boundary layer. Inside the boundary layer, 
we get from (\ref{4.4d}) the equation describing the behavior of pressure in the boundary layer in the form

\begin{equation}
\label{4.4e}
\frac{d P_0 }{ d\omega}=-4 \pi r_* \alpha_b\frac{P_0^2}{\dot M \Omega_{K \ast}(1-\omega)},\quad P_0=\rho_0 v_s^2,
\end{equation}
which has the solution

\begin{equation}
\label{4.4f}
\frac{1}{P_0}=-\frac{4 \pi r_* \alpha_b}{\dot M \Omega_{K \ast}}\ln(1-\omega)+D.
\end{equation}
Using the no-slip boundary condition,  $\omega=\omega_*$ on the star surface, let us express the integration constant $D$ via the
star surface pressure $P_*$,

\begin{equation}
\label{4.4g}
D=\frac{1}{P_*}+\frac{4 \pi r_* \alpha_b}{\dot M \Omega_{K \ast}}\ln(1-\omega_*).
\end{equation}
Then the solution has the following form:

\begin{equation}
\label{4.4h}
P_0=\biggl[\frac{1}{P_*}-\frac{4 \pi r_* \alpha_b}{\dot M \Omega_{K \ast}}\ln\left(\frac{1-\omega}{1-\omega_*}\right)\biggl]^{-1}.
\end{equation}
To calculate the pressure at the BL inner boundary $P_*$, we use the matching condition between the solution for the
boundary layer (\ref{4.4h}), and the disk (\ref{eq:11.P2.09}) at the boundary $\omega=\omega_b$, where $P_0=P_{0b}$:

\begin{equation}
\label{4.4i}
P_{0b}=\frac{2c \Omega_{K*} }{ 3 \alpha \varkappa_T}=
\biggl[\frac{1}{P_*}-\frac{4 \pi r_* \alpha_b}{\dot M 
\Omega_{K \ast}}\ln{\left(\frac{1-\omega_b}{1-\omega_*}
\right)}\biggl]^{-1}.
\end{equation}
From this, we obtain

\begin{equation}
\label{4.4j}
 P_*= \biggl[\frac{3 \alpha \varkappa_T }{ 2c  \Omega_{K*}}+\frac{4 \pi r_* \alpha_b}{\dot M \Omega_{K \ast}}\ln{\left(\frac{1-\omega_b}{1-\omega_*}\right)}\biggl]^{-1}.
\end{equation}

\begin{equation}
\label{4.4k}
 P_0= \biggl[\frac{3 \alpha \varkappa_T }{ 2c 
 \Omega_{K*}}-\frac{4 \pi r_* \alpha_b}{\dot M \Omega_{K 
 \ast}}\ln\left(\frac{1-\omega}{1-\omega_b}\right)
 \biggl]^{-1}.
\end{equation}
For the rotation speed of the neutron star approaching a critical value, BL thickness increases and coincides with
the disk thickness when
$\omega_*=\sqrt{1-\frac{v_{s*}}{r_*\Omega_{K \ast}}} \approx 1-\frac{v_{s*}}{2r_*\Omega_{K \ast}}$. Exceeding this angular velocity value of the neutron star
leads to the broadening and disappearance of the boundary layer. The above results hold for all equations of state
in accretion disks.

For radiation-dominated plasma with dominant radiation pressure, the temperature inside the boundary layer
$T_0$ and at its boundaries  $T_*,\,\,T_b$ is determined by the pressure

\begin{equation}
\label{4.4l}
T_0=\left(\frac{3 P_0}{a}\right)^{1/4},\qquad T_*=
\left(\frac{3 P_*}{a}\right)^{1/4},\qquad T_b=
\left(\frac{3 P_b}{a}\right)^{1/4}.
\end{equation}
Calculate approximately  the dependence of the vertical thickness of the boundary layer $h$ on $\omega$ using equation
\eqref{4.5}.

\begin{equation}
\label{4.5a}
\frac{d\omega }{ dr}
=\frac{\dot M \Omega_{K \ast}}{ 4 \pi r_* \alpha_b}\,
\frac{1-\omega}{\rho_0 v_s^2 H_b}.
\end{equation}
With account of the expression for  $H_b$ from (\ref{4.4d}), this equation can be written as

\begin{equation}
\label{4.5b}
\frac{d\omega }{ dr}
=\frac{\dot M \Omega_{K \ast}}{ 4 \pi  \alpha_b}\,
\frac{(1-\omega)(1-\omega^2)}{\rho_0 v_s^2 h^2};\qquad
[(1-\omega)(1-\omega^2)]^{-1}\frac{d\omega }{ dr}
=\frac{\dot M \Omega_{K \ast}}{ 4 \pi  \alpha_b}\,
\frac{1}{\rho_0 v_s^2 h^2}
\end{equation}
The half-thickness of the boundary layer is defined by the relation $h=\frac{v_s}{\Omega_{R*}}$. In the right-hand side of the second
equation, integration within the boundary layer is approximately replaced by multiplication by the quantity $H_b (1-\omega)$,
which, considering (\ref{4.4c}), gives

\begin{equation}
\label{4.5c}
\frac{1}{2}\frac{1}{1-\omega}+\frac{1}{4}\ln\frac{1+\omega}{1-\omega}
=\frac{\dot M \Omega_{K \ast}}{ 4 \pi  \alpha_b}\,
\frac{H_b(1-\omega)}{\rho_0 v_s^2 h^2} = \frac{\dot M }{ 4 \pi  
\alpha_b}\,\frac{1}{\rho_0 h^2 \Omega_{K \ast}r_\ast(1+
\omega)}.
\end{equation}
Neglecting the second, logarithmic term on the left-hand side, which is much smaller than the first term
throughout the entire interval $(1, \omega_\ast)$, the dependence 
$h^2(\omega)$ can be written in the form:

 \begin{equation}
\label{4.5d}
h^2
= \frac{\dot M }{ 4 \pi  
\alpha_b}\,\frac{2(1-\omega)}{\rho_0 \Omega_{K \ast}r_\ast(1+
\omega)}.
\end{equation}
For a radiation-dominated disk, at the matching point $\omega=\omega_b$, the disk thickness is given by the formula following
from \eqref{eq:11.P2.08}:

 \begin{equation}
\label{4.5e}
h_{db}\approx \frac{3}{8 \pi}\frac{\varkappa_T}{c}{\dot M}. 
\end{equation}
From the equality of the disk thicknesses (\ref{4.5d}) and (\ref{4.5e}) at
$\omega=\omega_b=\sqrt{1-\frac{v_{sb}}{r_*\Omega_{K \ast}}}\approx 1-\frac{v_{sb}}{2r_*\Omega_{K \ast}}$ , we get:

\begin{equation}
\label{4.5f}
\rho_0
=\frac{16\pi}{ 9\alpha_b \Omega_{K \ast}^2 r_*^2}\,
\frac{c^2}{\dot M  \varkappa_T^2}
\frac{v_{sb}}{1+ \omega_*}.
\end{equation} 
Assume, that the sound speed is continuous at the outer edge of the boundary layer and equal to the sound
speed at the inner edge of the disk, $v_{sb}=v_{sdb}$, where

  \begin{equation}
\label{4.5g}
v_{sdb}=\sqrt{\frac{P_{0db}}{\rho_{0db}}}= \frac{9\sqrt{\alpha}}{8\pi\sqrt{2}}\frac{\Omega_{K\ast}\dot M\varkappa_T}{c}. 
\end{equation}
Here, the expressions for pressure and density at the inner edge of the disk are taken from (\ref{eq:11.P2.09}) and 
(\ref{eq:11.P2.11}),
respectively. Substituting (\ref{4.5g}) into (\ref{4.5f}), we obtain the expression for the density in the boundary layer $\rho_0$ as:

\begin{equation}
\label{4.5h}
\rho_0
=\frac{2\sqrt{\alpha}}{\alpha_b\sqrt{2}} 
\frac{c}{\Omega_{K \ast} r_*^2 \varkappa_T}.
\end{equation} 
The density at the inner edge of the disk  $\rho_{0db}$ , as follows from (\ref{eq:11.P2.11}), has the form:

\begin{equation}
\label{4.5i}
\rho_{0db}=
=\frac{256 \pi^2}{27\alpha} 
\frac{c^3}{\Omega_{K \ast} \dot  M^2 \varkappa_T^3}.
\end{equation}
For the ratio of the two density expressions at the edge of the boundary layer, we get: 
 
\begin{equation}
\label{4.5j}
\frac{\rho_{0}}{\rho_{0db}}
=\frac{27}{128\pi^2}\frac{\alpha^{3/2}}{\alpha_b \sqrt 2} 
\left(\frac{\varkappa_T \dot M}{c r_*}\right)^2.
\end{equation}
For estimation purposes, it is more convenient to express this ratio as a function of the stellar luminosity $L$
due to accretion, rather than in terms of the mass accretion rate
$\dot M$. These values are uniquely related under the
assumption of constant accretion efficiency  $\eta$, such that:

 \begin{equation}
\label{4.5k}
L=\eta \dot M c^2.
\end{equation}} 
It is also convenient to express luminosity in units of the critical Eddington luminosity $L_c$, and to express
the stellar radius in units of the Schwarzschild gravitational radius $r_g$:

\begin{equation}
\label{4.5l}
L_c=\frac{4\pi c GM}{\varkappa_T},\qquad r_g=\frac{2GM}{c^2}.
\end{equation}} 
As a result, we obtain the density ratio in the form:

\begin{equation}
\label{4.5m}
\frac{\rho_{0}}{\rho_{0db}}
=\frac{1}{\eta^2}\frac{27}{128\pi^2}\frac{\alpha^{3/2}}{\alpha_b \sqrt 2} 
\left(\frac{r_g}{2r_*}\right)^2 \left(\frac{L}{L_c}\right)^2.
\end{equation}
We also take into account that the radii of most neutron stars are approximately $r_*\approx 3r_g$ . Then we obtain, for $\eta=0.1$:

\begin{equation}
\label{4.5n}
\frac{\rho_{0}}{\rho_{0db}}
=\frac{1}{\eta^2}\frac{27}{36\cdot128\pi^2}\frac{\alpha^{3/2}}{\alpha_b \sqrt 2} \approx \frac{3\cdot 10^{-4}}{\eta^2}\frac{\alpha^{3/2}}{\alpha_b \sqrt 2}
 \left(\frac{L}{L_c}\right)^2=0.03\frac{\alpha^{3/2}}{\alpha_b \sqrt 2}
 \left(\frac{L}{L_c}\right)^2
\end{equation}
As we see from the above, the value $\rho_0$ is assumed to be constant within the boundary layer and corresponds
to the average density inside the boundary layer. Therefore, its smallness compared to the density in the disk may
be attributed to the decrease in density within the boundary layer due to rising temperature and significant vertical
expansion. As the accretion disk luminosity approaches the critical value, the vertical expansion of the boundary layer
is expected to form a collimated jet perpendicular to the disk plane (Fig. 1). To investigate such jets, numerical
simulations of the boundary layer around a neutron star in two- or three-dimensional models are required. Such models
are too complex for current computational capabilities so far.

\begin{figure}
   		\includegraphics[width=0.99\textwidth]{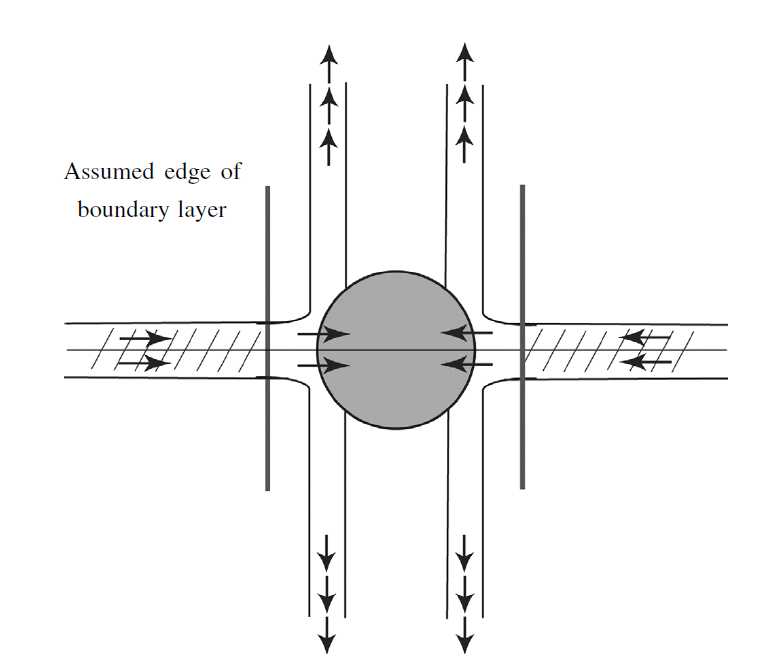}	
   	\caption{
 Qualitative picture of accretion onto a neutron star with a
weak magnetic field in the boundary layer formation region. In
the interior part of the disk, the plasma is assumed to be radiation dominated,
which corresponds to high luminosity during
accretion. The arrows indicate the direction of gas motion within
the boundary layer, where part of the matter is incorporated into
the neutron star. At high luminosities, the radiation force
accelerates part of the plasma in the vertical direction, leading to the formation of collimated jets. The assumed edge of the
boundary layer is located where the characteristic BL thickness,
as given by equation (\ref{4.4c}), becomes comparable to the vertical
thickness of the disk.   	
}
   	\label{bl}
   \end{figure}	

\section{Conclusion}

The significant simplifications and approximations employed in this work have allowed us to construct a semiqualitative
model of the boundary layer around a rotating neutron star (NS) with a weak magnetic field that does
not affect the dynamics of the boundary layer. We considered the case of strong accretion onto the NS, where radiation
pressure dominates in the interior part of the disk and boundary layer. A simple derivation was done for the total
luminosity of the boundary layer (24) as a function of the neutron star angular velocity.
An explicit formula (44) was obtained for the dependence of the vertical thickness of the boundary layer on
radius (or angular velocity), showing that it increases as $h\sim \sqrt{(1-\omega)/(1+\omega)},\,\,\omega \geq \omega_*$ when approaching the NS, and
remains continuous at the outer edge of the accretion disk boundary layer, see \eqref{4.5d} and \eqref{4.5e}. Formulas \eqref{4.4k} and \eqref{4.4l}
determine the pressure and temperature distributions within the boundary layer, respectively. The provided equations
illustrate a continuous transition of parameters across the area where the boundary layer of the accretion disk begins.
To formally localize the boundary between the disk and the BL, we defined the point at which the radial
thickness of the boundary layer becomes equal to its vertical thickness. The boundary layer in this case lies between
this point and the surface of the neutron star. A similar criterion can be applied to determine the state of the star with
an accretion disk in which the boundary layer disappears as the angular velocity of the star increases. The NS rotation
rate in this state is close to the critical value, and falls short of it by a small but finite amount.
The model of the boundary layer developed here appears to be qualitatively valid, despite several crude
approximations and simplifications made in its construction. This is evidenced by the fact that, when applying it to
a star with critical Eddington luminosity, all the derived parameters are consistent with general expectations of the
model, provided that we use independent estimates for the neutron star mass and the accretion efficiency $\eta$.


\begin{thebibliography}{99}

 \bibitem{sht85}
 S. Shapiro, S. Teukolsky, Black Holes, White Dwarfs, and Neutron Stars, Moscow, Mir (1985).

 \bibitem{lip87}
V.M. Lipunov, Astrophysics of Neutron Stars, Moscow, Nauka (1987).

 \bibitem{sha}
N.I. Shakura, Astron. Zh., 49, 921 (1972).
(1973, Sov. Astron., {\bf 16}, 756)

\bibitem{shs}
N.I. Shakura, R.A. Sunyaev, Astron. Astrophys., 24, 337 (1973).

\bibitem{nt}
Novikov, I.D., Thorne, K.S., 1973. In: DeWitt, C., DeWitt, B.
(Eds.), Black Holes. Gordon and Breach, New York, p. 345

\bibitem{bkb}
Bisnovatyi-Kogan, G.S., Blinnikov S.I. 1977, A\&A, {\bf 59}, 111

\bibitem{pbk}
Paczyn'ski, B., Bisnovatyi-Kogan, G.S., 1981. AcA {\bf 31}, 283

\bibitem{bkfr}
G.S. Bisnovatyi-Kogan, A.M. Fridman, Astron. Zh., 46, 721 (1969). (1970, Sov. Astron., {\bf 13}, 566)

\bibitem{pring1} Pringle J., 1977. MNRAS {\bf 178}, 195

\bibitem{pring2} Pringle J., Savonije G.,  1979. MNRAS {\bf 187}, 777
 
\bibitem{tyl1} Tylenda, R. 1977. Acta Astron., {\bf  27}, 235

\bibitem{tyl2} Tylenda, R. 1981. Acta Astron., {\bf 31},  267

\bibitem{reg}
Regev O. 1983, Astron. Ap., {\bf 126}, 146 

\bibitem{pap} Papaloizou, J.C.B., Stanley G.Q.G., 1986.  MNRAS,{\bf 220},  593

\bibitem{ss}
Shakura, N.I., Sunyaev, R.A., 1988, Adv. Space Res. {\bf 8} (2), 135

\bibitem{pop}
Popham P. \&  Narayan R. 1995, ApJ,  {\bf 442}, 337 
 
\bibitem{is}N.A. Inogamov, R.A. Sunyaev, Pisma v Astron. Zh., 25, 323 (1999). (1999, Astron. Lett.,{\bf 25}, 269) 

\bibitem{bk2010}
Bisnovatyi-Kogan G.S.
"Stellar Physics 2: Stellar evolution and stability"
Springer. 2010


\bibitem{bkl2001}
Bisnovatyi-Kogan G.S., Lovelace, R.V.E. 2001, New Astronomy Reviews
{\bf 45}, 663


\bibitem{nay81}
Nayfet A., 1981, Introduction into Perturbation Techniques.
John Wiley \& Sons, New York

\bibitem{bk94}
Bisnovatyi-Kogan G.S. 1994,
MNRAS {\bf 269}, 557


\bibitem{bk93}
Bisnovatyi-Kogan G.S. 1993, A\&A {\bf 274}, 796


\end{thebibliography}
\end{document}